\documentclass[12pt, onecolumn]{IEEEtran}

\makeatletter
\def\ps@headings{%
\def\@oddhead{\mbox{}\scriptsize\rightmark \hfil \thepage}%
\def\@evenhead{\scriptsize\thepage \hfil \leftmark\mbox{}}%
\def\@oddfoot{}%
\def\@evenfoot{}}
\makeatother 

\usepackage{graphicx}
\usepackage{amsmath}
\usepackage{amsmath, amsfonts,epsfig, multirow, floatflt}
\usepackage{epsfig,graphics,graphicx,latexsym,amsfonts,amssymb,amsmath,verbatim}
\usepackage{subfigure}
\usepackage{cite}
\usepackage{xcolor}
\usepackage{soul}
\usepackage{color}
\graphicspath{{figures/}}

\begin{document}
\baselineskip 24pt
\parskip 9pt
\thispagestyle{empty}

\baselineskip 19pt
\parskip 4pt

\setcounter{page}{1}

\begin{center}
\vspace*{0mm}

{\LARGE \bf Radio Resource Allocation in LTE-Advanced Cellular
Networks with M2M Communications} \vspace*{10mm}

{\normalsize Kan Zheng$^\ast$, {\it Senior Member, IEEE}, Fanglong Hu$^\ast$, Wei Xiang$^\dag$, {\it Senior Member, IEEE}, Mischa Dohler, {\it Senior Member, IEEE}$^\ddag$ and Wenbo Wang$^\ast$, {\it Member, IEEE}\\

\vspace{0.3cm}

\small $^\ast$Wireless Signal Processing and Network Lab\\
Key laboratory of Universal Wireless Communication, Ministry of Education \\
Beijing University of Posts \& Telecommunications\\
Beijing, China, 100088 \\

\small $^\dag$ Faculty of Engineering and Surveying \\
University of Southern Queensland \\
Toowoomba, QLD 4350, Australia \\

$^\ddag$ Centre Tecnologic de Telecommunications de Catalunya (CTTC) \\
Barcelona, Spain, 08860 \\

E-mail: \tt{kzheng@ieee.org} \\
}
\end{center}

\vspace*{10mm}

\begin{center} {\bf Abstract}
\end{center}
Machine-to-machine (M2M) communications are expected to provide
ubiquitous connectivity between machines without the need of human
intervention. To support such a large number of autonomous devices,
the M2M system architecture needs to be extremely power and
spectrally efficient. This article thus briefly reviews the features
of M2M services in the third generation (3G) long-term evolution and
its advancement (LTE-Advanced) networks. Architectural enhancements
are then presented for supporting M2M services in LTE-Advanced
cellular networks. To increase spectral efficiency, the same
spectrum is expected to be utilized for human-to-human (H2H)
communications as well as M2M communications. We therefore present
various radio resource allocation schemes and quantify their utility
in LTE-Advanced cellular networks. System-level simulation results
are provided to validate the performance effectiveness of M2M
communications in LTE-Advanced cellular networks. \par

\begin{flushleft}
\end{flushleft}
\newpage
\baselineskip 24pt
\parskip 9pt

\clearpage
\section{Introduction}

Machine-to-machine (M2M) communication is concerned with connecting
communication-enabled devices in an unprecedented way, thus enabling in parts the Internet of Things (IoT)~\cite{IoT,IoT1}. With M2M communications, devices ``talk" to each
other through wired or wireless connections and share data without
direct human intervention. The use of M2M communications is
particularly well suited to interact with a large number of remote
devices acting as the interface with end customers, utilities, etc. In this way,
devices such as smart meters, signboards, cameras, remote sensors,
laptops, and appliances can be interconnected to support a variety of new applications~\cite{M2Mapplications}.
\par

With the rapid development of the third generation (3G) long-term
evolution cellular networks and its advancement (LTE-Advanced), M2M
communications via LTE-Advanced cellular networks with widespread
coverage is expected to constitute a significant part of the IoT.
For mobile service operators, services through M2M communications
have promising and strategic values. For instance, a large number of
M2M services are non-real time and typically consume little bandwidth, with
minimal impact on the capacity of radio access networks (RANs). With
M2M services supported by information and communication technologies
(ICT), operators can expand their end-to-end information solutions into
industries beyond the currently supported. \par

%Thus, ICT can penetrate into end systems
%of production and operation in diversified industries, leading to effective integration of ICT with industrialization.\par

Unlike traditional human-to-human (H2H) services, such as voice and
web streaming, M2M services often have very different requirements
on a communication system due to their specific
features~\cite{3gpp368}. Another distinguishing characteristic in
cellular networks with M2M communication is the large increase in
the number of machine-type communication (MTC) devices. Both of them
bring forth new challenges for LTE-Advanced cellular networks,
demanding significant improvements in the efficiency of radio
resource utilization. Meanwhile, new M2M services have to have
little or even no impact on existing H2H services in cellular
networks~\cite{3gpp888}. Preliminary studies on M2M communications
have primarily thus focused on service requirements, the functional
architecture and applications~\cite{3gpp368,ETSIM2MService}.\par

Concerning the service requirements, M2M applications are quite
different from their H2H counterparts since M2M services have their
own very unique characteristics~\cite{3gpp368,Mischa_tutorial}.
Moreover, Quality-of-Service (QoS) requirements of different types
of M2M services vary widely and are reflected in the MTC service
features: group-based communications, low or no mobility,
time-controlled, time-tolerant, small data transmission, secure
connection, MTC monitoring, priority alarm messages, extra low power
consumption, etc. These service requirements then dictate the
architectural design, to be discussed below.\par

With an architecture in place, numerous challenges remain for radio
resource management (RRM)  for M2M communication in LTE-Advanced
cellular networks. For example, time and frequency resources are to
be shared between H2H users and MTC devices (MTCDs), thus resulting
in co-channel interference among them~\cite{D2D}. Such co-channel
interference plays a detrimental role in degrading the performance
of the LTE-Advanced cellular networks with M2M communications.
Furthermore, differentiated QoS requirements between H2H users and
MTCDs have to be accommodated, which requires different interference
tolerances for different types of users and devices. To the best of
the authors' knowledge, there has been no work in the literature on
radio resource allocation for LTE-Advanced cellular networks with
M2M communications so far.\par

The scope of this article is hence to examine how H2H users and MTC
devices can share available radio resources efficiently so as to mitigate co-channel interference and thus enhance network
efficiency. We first introduce some architectural enhancements needed to fulfill above MTC
service requirements. Then, several radio resource allocation schemes are
proposed for LTE-Advanced cellular networks with M2M communications.
We then analyze and assess their performance extensively through
system-level simulations and study H2H-M2M coexistence issues.\par

\section{Architectural Enhancements to Fulfill M2M Service Requirements}

In order to meet the requirements of LTE-Advanced such as peak data
rates of up to 1 Gbit/s, more spectrum bands are needed. Besides the
existing carriers for 3G networks, spectrum bands located at 450-470
MHz, 698-790 MHz, 2.3-2.4 GHz and 3.4-3.6 GHz can be used for the
deployment of LTE and LTE-Advanced networks \cite{M1645}. Moreover,
LTE-Advanced has been defined to support scalable carrier bandwidth
exceeding 20 MHz, potentially up to 100 MHz, in a variety of
carriers for deployments. \par

The current RAN for LTE-Advanced consists of a single node, i.e.,
the eNodeB (eNB) that provides the user plane and control plane
protocol terminations towards the user equipment (UE).  It is a fully
distributed radio access network architecture, where eNBs may be
interconnected with each other by means of the X2 interface.
Meanwhile, eNBs are connected through the S1 interface to the core
network. In each eNB, there exist the PHYsical (PHY), Medium Access
Control (MAC), Radio Link Control (RLC), and Packet Data Control
Protocol (PDCP) layers that implement the functionality of
user-plane header-compression and encryption. The current 3G LTE
cellular network is designed only for providing H2H services for
user equipments (UEs). However, with the introduction of M2M
communications, the network architecture needs to be improved to
accommodate M2M service requirements without sacrificing the
qualities of current H2H services.\par

\subsection{MTCD-Related Communications}

\begin{figure}[thbp]
\begin{center}
\subfigure[Direct
transmission.]{\includegraphics[width=5in]{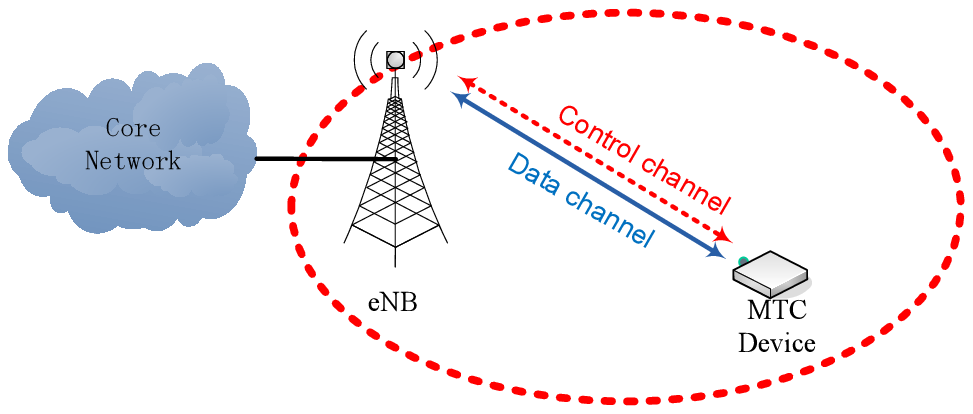}
\label{fig_Directlink} } \\
\subfigure[Multi-hop transmission.]{\includegraphics[width=5in]{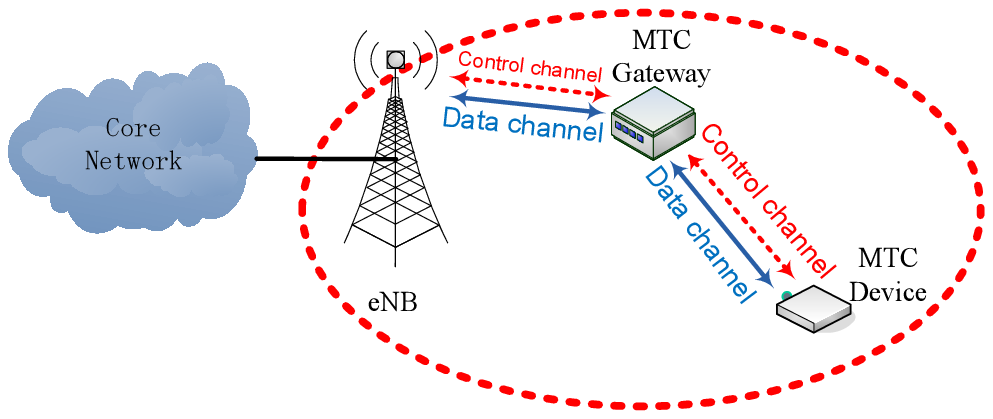}\label{fig_Relaylink}}
\\  \subfigure[Peer-to-peer transmission.]{\includegraphics[width=5in]{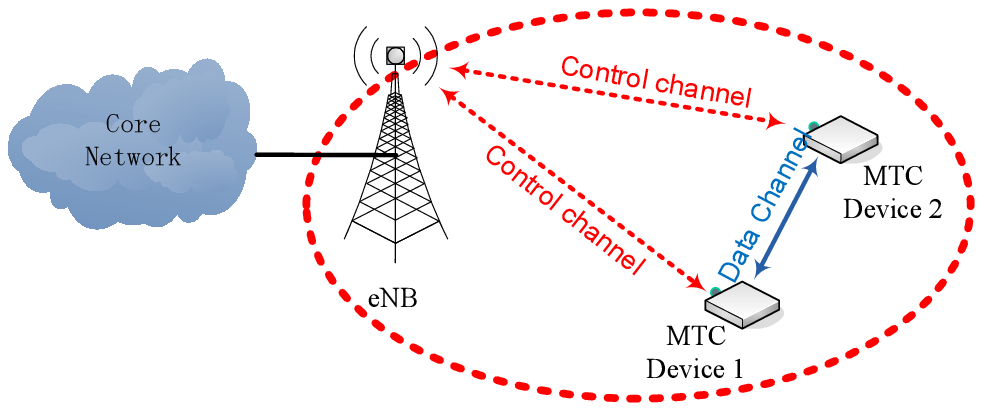}\label{fig_Peer2peerlink}}
\caption{Illustration of MTCD-related transmission.}
\label{fig_m2m_communication}
\end{center}
\end{figure}

To enable M2M communications, two new network elements, i.e., the
MTCD and MTC gateway (MTCG), appear in LTE-Advanced cellular
networks. A MTCD is a user equipment (UE) designed for
machine-type communications, which communicates through a cellular
network with an MTC server and/or other MTCDs. The network requires
an MTCG gateway to facilitate communications among a great many MTCDs and
to provide a connection to a backhaul that reaches the Internet. The
MTCG will be able to intelligently manage power consumptions of the
network, and provide an efficient path for communications between
MTCDs. Three different M2M communications methods are feasible, as
illustrated in Fig.~\ref{fig_m2m_communication}. \par

\subsubsection{Direct transmission between MTCD and eNB}
Similar to a normal UE, an MTCD has the ability to establish a
direct link with its donor eNB. Therefore, there exist strong
similarities between the eNB-to-UE and eNB-to-MTCD links. On the
other hand, MTCDs normally appear in large quantities in the M2M
networks and thus exhibit the service feature of group-based
communications. In certain time instants, intense competition for
radio resources may occur. For instance, one or more MTC groups send
communication requests to an eNB simultaneously, which may cause
network congestion, resulting in performance degradation for both
M2M and H2H services. Therefore, additional efforts have to be made
to tackle such kinds of problems, when a large quantity of MTCDs
communicate with the eNB directly.
\par

\subsubsection{Multi-hop transmission with the aid of an MTC gateway}
In order to mitigate or eliminate negative effects of M2M communications
on H2H communications, an MTC gateway can be deployed in celluar
networks, where all MTCDs are connected to the eNB indirectly through the relaying of the MTCG. In other words, the end-to-end
communication between the eNB and MTCDs may occur via more than one hop,
e.g., the eNB-to-MTCG and MTCG-to-MTCD links. Besides, MTCDs may
establish peer-to-peer communications with each other with the
aid of the MTCG or eNB. The eNB-to-MTCG wireless link is based on 3G
LTE specifications, whereas the MTCG-to-MTCD and MTCD-to-MTCD
communications can  either be via 3G LTE specifications or other
wireless communications protocols such as IEEE 802.15.x. The resulting
multi-level network management problem can be handled with the aid
of the MTCG. Each MTCD is controlled by its donor MTCG, which is
managed by the eNB. The introduction of the MTCG makes the network
topology more complex, leading to challenges as well as
opportunities.

MTCDs are usually grouped for control, management or charging facilities. The MTCDs within
the same group can be in the same area and/or possess the same MTC
features. Each MTCG can serve one or more groups.

\subsubsection{Peer-to-peer transmission between MTCDs}
An MTCD may communicate locally with other entities, which provide
the MTCD with raw data for processing and communicating to the MTC server
and/or other MTCDs. Compared to other local connectivity solutions,
such as IEEE 802.11a or IEEE 802.15.x, peer-to-peer transmission between MTCDs
supported by a cellular network offers appealing advantages. The
cellular network can broadcast local services available within a much
wider coverage area. Thus, for automated service discovery, the MTCDs do
not have to constantly scan for available local access points (APs)
as in case of IEEE 802.11a. This is advantageous since leading to significantly
reduced power consumption for scanning. With the knowledge of
encryption keys at both MTCDs involved in peer-to-peer
communications, a secure connection can be established without manual
pairing of devices or entering encryption keys. Moreover, through the control of the eNB via peer-to-peer communications, the
interference to other cellular receivers can be limited or
mitigated. \par

\subsection{Architectural Enhancements}

%The radio access network of cellular networks typically adopts the
%star topology, e.g., an eNB is located in the center of a cell while
%all UEs communicate with the eNB directly.

\begin{figure}[thbp]
\begin{center}
\includegraphics[width=6in]{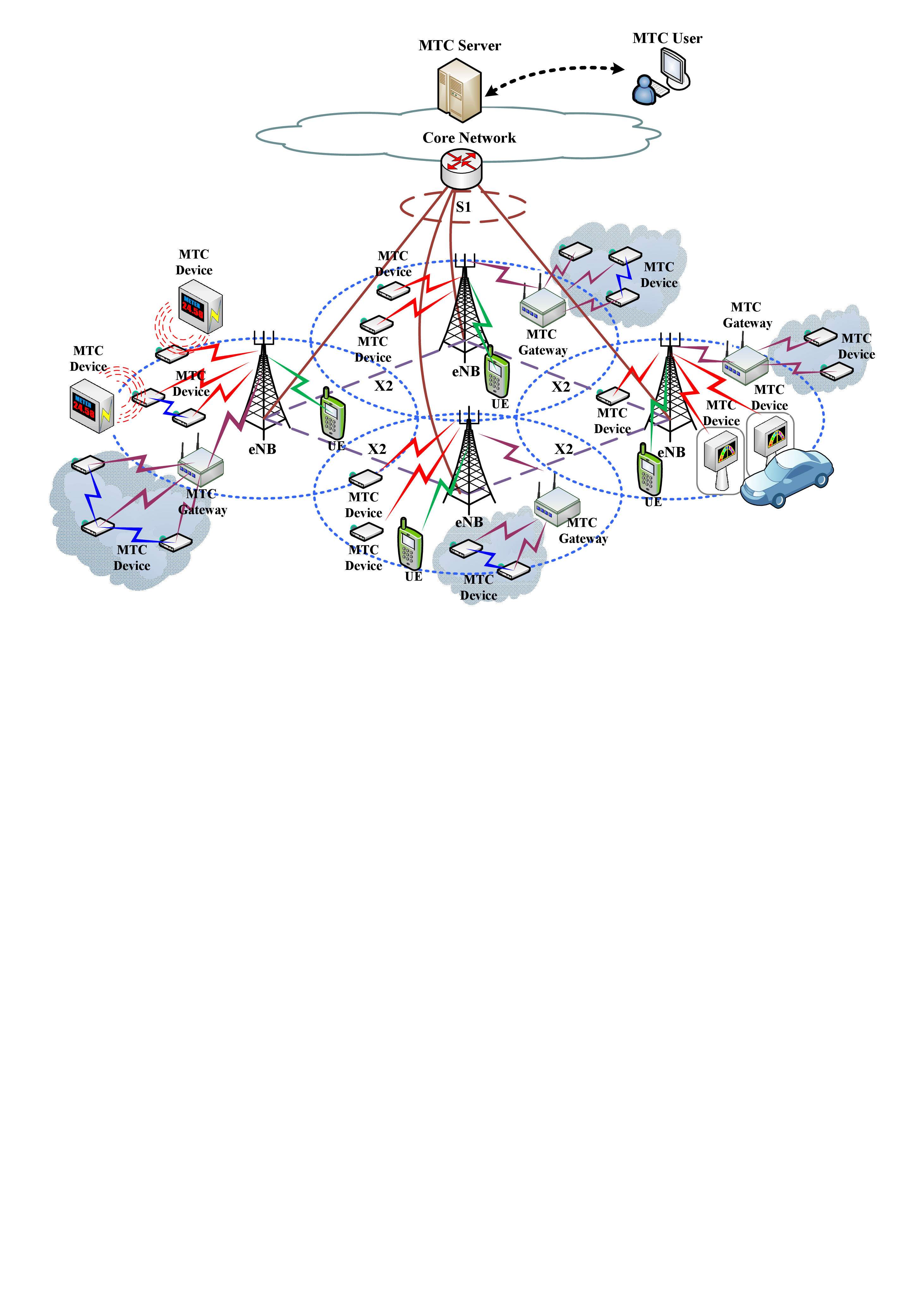}
\caption{Architectural enhancements to LTE-Advanced cellular
networks with M2M communications.} \label{fig_m2m_architecture}
\end{center}
\end{figure}

In order to support M2M communications, the RAN architecture needs
to be enhanced to enable coexisting communications between
MTCD-related and H2H communications in LTE-Advanced cellular
networks. Fig.~\ref{fig_m2m_architecture} gives an example of
architectural enhancements to the M2M cellular network. \par

Apart from direct transmission, MTCDs can also establish communications with their donor eNBs through multi-hop transmission.
To avoid self-interference and reduce implementation complexity, half-duplex MTCGs are preferred for deployment in the networks. Furthermore, when local services are available
between nearby MTCDs, peer-to-peer communications provided by
cellular networks may appear to be a local connectivity solution.
\par

In such fairly intricate M2M cellular networks, how to assign and coordinate radio
resources to different classes of transmissions becomes a critical issue,
which will be dealt with in the next section.
\par

\section{Radio resource allocation for M2M Communications}

Introducing M2M communications to LTE-Advanced cellular networks
should not come at the expense of significantly degraded performance
for existing H2H communications. There are two major methods for
radio resource allocation between M2M and H2H communications, i.e.,
orthogonal and shared resource allocation. Collocating M2M and H2H
communications in orthogonal channels is a simple solution but leads
to low spectral efficiency from a system level perspective. To
achieve higher spectral efficiency, M2M communications can reuse the
radio resources assigned to H2H communications, resulting in shared
channel allocation. However, this will cause an increased level of
interference in comparison with orthogonal channel allocation. \par

\begin{figure}[thbp]
\begin{center}
\subfigure[Transmission
links.]{\includegraphics[width=4.5in]{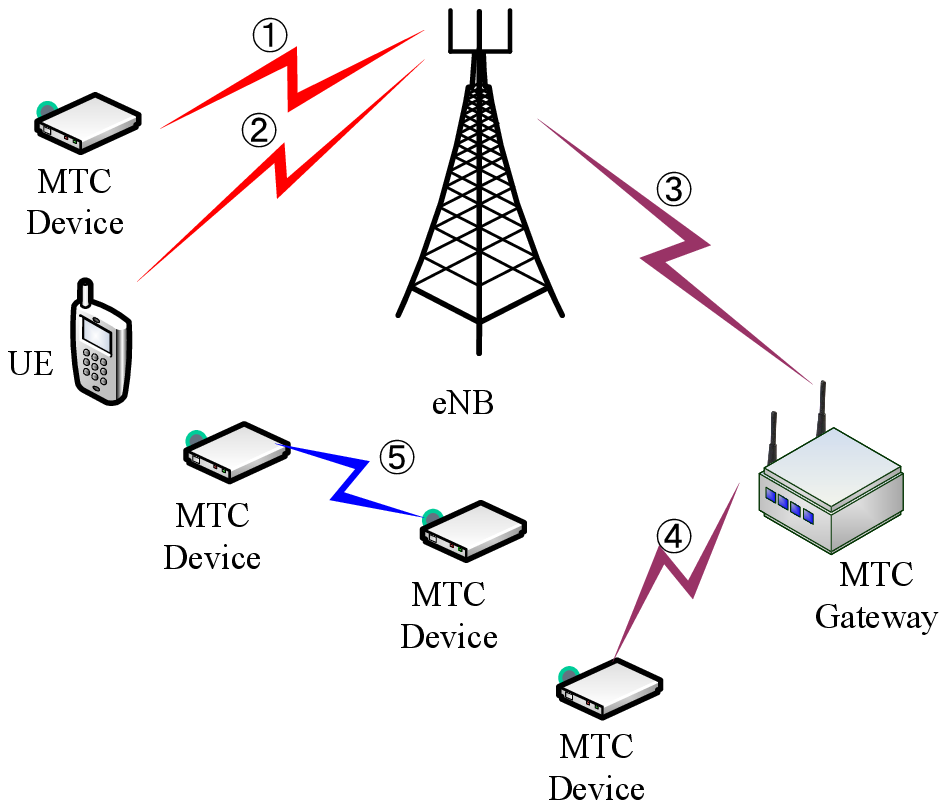} \label{fig_links}}
\subfigure[Example of radio resource
partition.]{\includegraphics[width=4.5in]{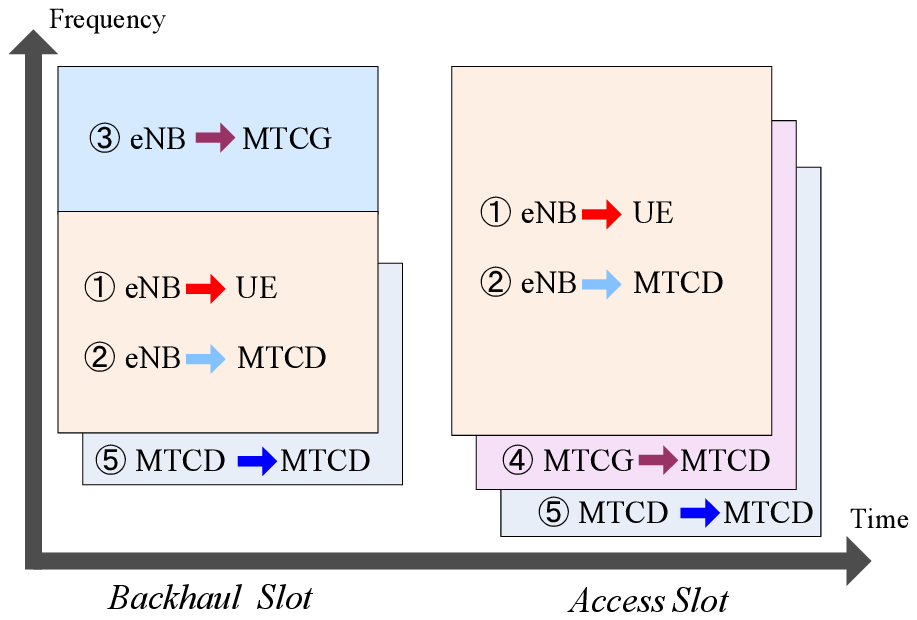}\label{fig_partition}}
\caption{Illustration of radio resource partition in LTE-Advanced
networks with M2M communication.} \label{fig_radioresource}
\end{center}
\end{figure}

In LTE-Advanced networks, radio resources are usually divided into
resource blocks (RBs) along the frequency domain per time slot,
which is also referred to as subchannels in radio resource
allocation. In mixed H2H and M2M communications networks, there are
usually five types of links  as illustrated in Fig.~\ref{fig_links};
namely, 1) the eNB-to-UE link; 2) the eNB-to-MTCD link; 3) the
eNB-to-MTCG link; 4) the MTCG-to-MTCD link; and 5) the MTCD-to-MTCD
link. When radio resources are shared among these links,
interference becomes a challenging issue. Therefore, it is essential
to first design the efficient radio resource partition in such
networks.
\par

The radio resource partition aims at applying restrictions to the
radio resource management in a coordinated way among nodes. These
restrictions can be either on the available radio resources or in
the form of restrictions on the transmit power that can be applied
to certain radio resources. Such restrictions provide the
possibility for improvement in Signal-to-Interference-plus-Noise
ratio (SINR), and consequently to the cell edge performance and
coverage. It is critical to exploit the characteristics of links to
obtain the well-designed coordination pattern, which can achieve
performance gain for almost all users in the network.
Fig.~\ref{fig_partition} shows an example of radio resource
partition pattern for the downlink transmission of LTE-Advanced
cellular networks with M2M communications. On the assumption of
half-duplex MTCGs, every two time slots are grouped together as one
basic unit for transmission. In the first slot, termed as the
\emph{backhaul slot}, the MTCGs receive signals from the eNB. In the
second slot, called the \emph{access slot}, MTCGs send the data to
their serving MTCDs.\par

In the \emph{backhaul slot}, the eNB-to-MTCG transmission link
has to be reliable to ensure the service quality
of the MTCDs associated with the MTCG. Thus, the eNB-to-MTCG links are assigned orthogonal parts of the radio resources, whereas all other links directly associated with the eNB share the channel. In the \emph{access slot}, all links except for the eNB-to-MTCG links share all the radio resources using various methods.
\par

MTCD-to-MTCD communications are envisaged to take place only locally
with relatively low power, and using either uplink or downlink
channel. This implies that these links do not interfere with any
other links, whereas the inverse of course does not hold true. The
MTCG-to-MTCD link, however, can strictly generate interference with
ongoing communication links. However, since these devices typically
serve some spatially very small areas where coverage is typically
poor, the impact onto the other links is neglected here.
Incorporating a complete interference scenario is possible but
unlikely to change the design insides.
\par

With the resource partition pattern given in
Fig.~\ref{fig_partition}, not all the interferences between
different links have to be specifically dealt with. So, we only
focus on resource allocation between some of links in the following
parts. \par

\subsection{Orthogonal Allocation for the eNB-to-MTCG Link}

Radio resource allocation and scheduling between the MTCG and
MTCDs can be carried out at the MTCG in coordination with its donor
eNB. Instead of communicating with an eNB directly, the MTCDs associated
with a MTCG first establish a link with the MTCG. Via such multi-hop
transmission, intense competition against radio resources can be
mitigated especially when enormous MTCDs request access to the network resources simultaneously. In addition, the radio resources can also be reused
between MTCGs in the case of multiple MTCGs per cell to improve on the spectral efficiency of the network.
\par

As mentioned before, there is no co-channel interference between the
eNB-to-MTCG link and other links due to orthogonal channel
allocation in the \emph{backhaul slot}. For the purpose of achieving
high spectral efficiency, resource allocation for the eNB-to-MTCG
link needs to be adjusted semi-statically according to service
demands from the MTCDs associated with the MTCG. If there are not
enough resources available for data transmission between the eNB and
MTCG, the associated MTCDs can not be served in time, resulting in
QoS degradation. Otherwise, when excessive resources are assigned to
the eNB-to-MTCG link, the QoS performance of other users such as UEs
and other MTCDs may suffer due to insufficient radio resources.
Hence, resource partition in the backhaul slot is rather crucial to
the overall system performance.\par

Fortunately, MTCDs usually possess common service features of
\emph{small data transmission} and \emph{time-tolerance}, which
implies that the average or maximum data rate of MTCDs can be easily
known according to the service type or device type. With the
knowledge of M2M services, the eNB can first roughly estimate the
total data rate of all MTCDs attached to a MTCG. Then, the number of
RBs needed for the transmission between the eNB and MTCGs in the
backhaul slot can approximately be calculated through dividing the
total data rate by the average data rate per RB in the eNB-to-MTCG
link.
\par

\subsection{Scheduling Between the eNB-to-UE and eNB-to-MTCD Links}

%In the eNB-UE and eNB-MTCD links, the orthogonality between them is
%necessary to be kept since the eNB usually cannot transmit different
%signals to more than one destination on the same resource. In order
%to well support M2M transmission and not degrade the performance of
%the H2H transmission, utility based resource allocation is
%considered.
%
%\par

For LTE-Advanced cellular networks with both H2H and M2M
services, the user utility of a service is more informative than the simple QoS indicator
due to the diversity of the applications. Generally speaking, the user
utility of a service is a measurement of its QoS performance
based on the provided network services such as the bandwidth,
transmission delay and loss ratio. It describes the satisfaction
level of the service delivered to the application. Here we focus on
the user utility as a function of the achievable data rate only,
which is most commonly used in the literature. With the
control of the eNB, radio resources can be efficiently shared
between the eNB-to-UE and eNB-to-MTCD links by using the
utility-based scheduling scheme.\par

We can classify network applications into four classes and
their features are shown as follows~\cite{utilityfunctions}:
\begin{itemize}
\item \emph{Class 1 (Elastic Applications)}:
Such applications are rather tolerant of delays. Traditional H2H
data applications like file transfer, electronic mail are typical
ones of this kind.  Another example is file downloading of
remote MTCDs from MTC servers. Their user utility has diminishing marginal improvements with incremental increase in the achievable data rate, and can be described as a strict concave function.

\item \emph{Class 2 (Hard Real-Time Applications)}: These applications need their
data to be served within a given delay constraint. Otherwise, there
is no extra utility gain even with further increase of the data
rate. An example of such applications is traditional telephony. For
applications with hard real-time requirements, the user utility is a
step function of the achievable data rate. Vehicle and asset tracking, a typical M2M
service application,  has to monitor and manage
the MTCDs in real-time, which is also a hard real-time application.\par

\item \emph{Class 3 (Delay-Adaptive Applications)}:
Applications like audio and video services are delay sensitive.
However, most of these applications can be made rather tolerant of
occasional delay-bound violation and dropped packets. They have an
intrinsic data rate requirement, and the user utility deteriorates
rapidly only when the achievable data rate is below the requirement.
There are a great number of such applications in M2M communications, e.g.,
remote monitoring in e-Health services. \par

\item \emph{Class 4 (Rate-Adaptive Applications)}:
Rate-adaptive applications adjust their transmission rates according
to available radio resources while maintaining moderate delays. Thus, the
performance of these applications highly depends on the scheduling
scheme and the quality of the underlying wireless channel.
Obviously, the increase of utility with the increase of the data
rate is only marginal at high data rates. Conversely, the increase
of utility will not be significant at very low data rates owing to
the unbearably low signal quality.
\end{itemize}

\begin{figure}[thbp]
\begin{center}
\subfigure[Class
1.]{\includegraphics[width=3.5in]{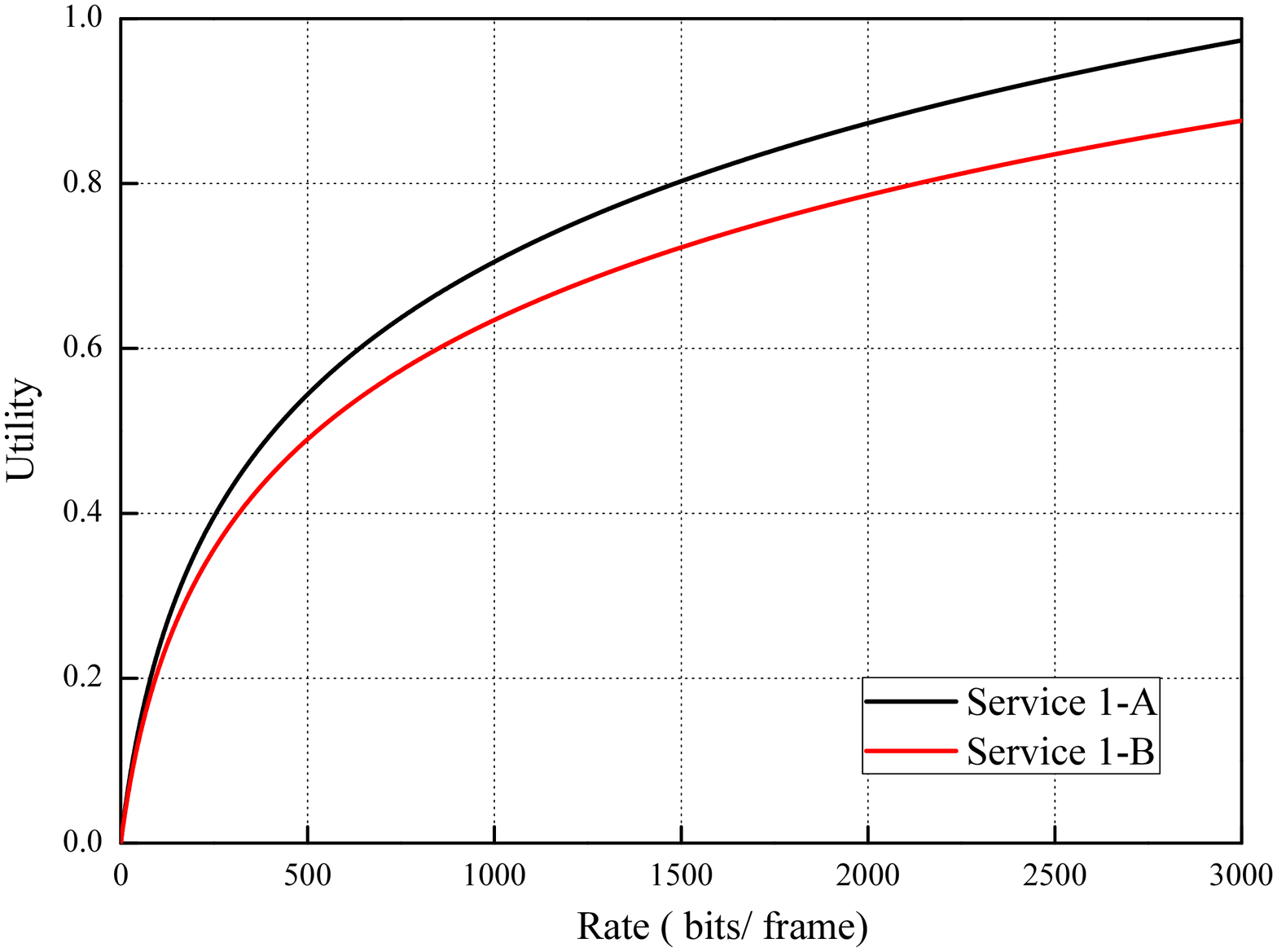}
\label{fig_Utility_Class1} } \subfigure[Class
2.]{\includegraphics[width=3.5in]{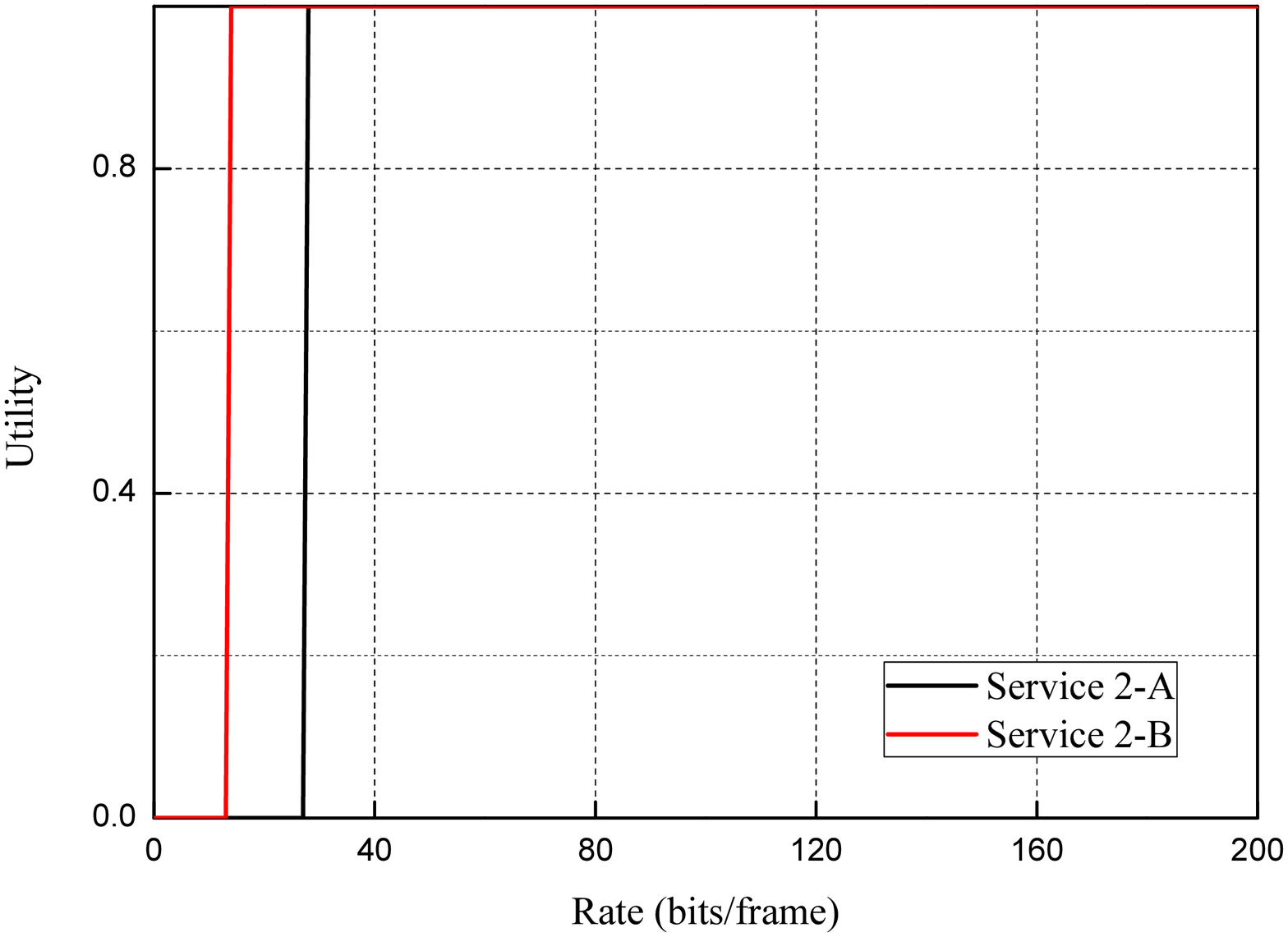}\label{fig_Utility_Class2}}
\\  \subfigure[Class 3.]{\includegraphics[width=3.5in]{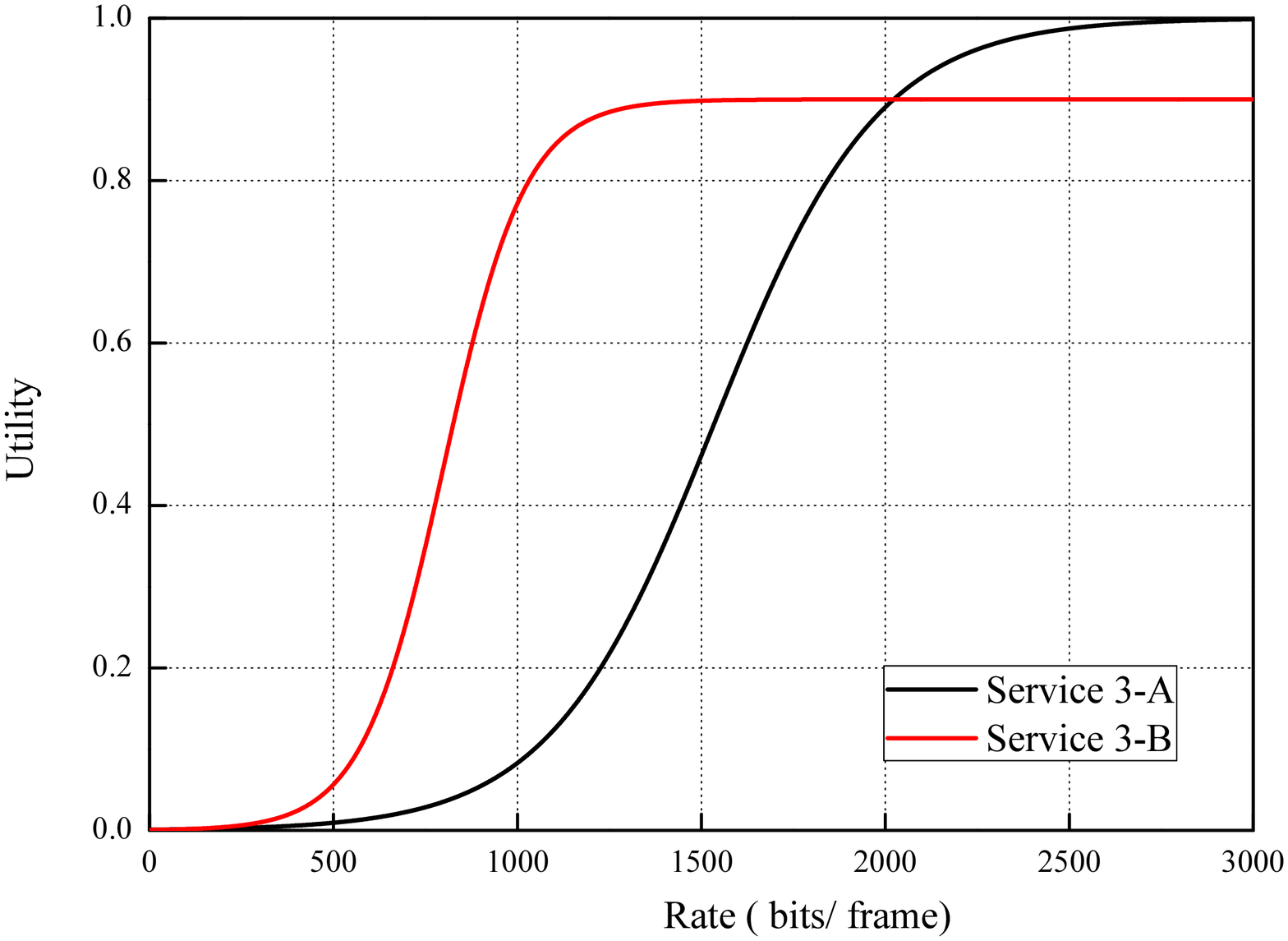}\label{fig_Utility_Class3}}
\subfigure[Class
4.]{\includegraphics[width=3.5in]{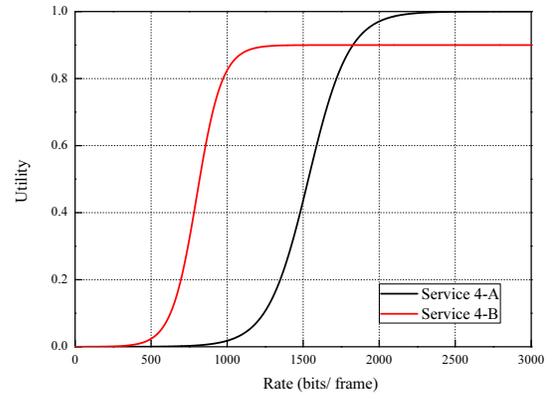}\label{fig_Utility_Class4}}
\caption{Example utility functions for four different classes of
applications.} \label{fig_Utility_Functions}
\end{center}
\end{figure}

\noindent Fig.~\ref{fig_Utility_Functions} illustrates four example
utility functions corresponding to the four classes discussed above.
Due to different requirements of the applications, it is likely that
H2H and M2M services have various formats or parameters for their
utility functions, even though they fall under the same class.\par

Usually the resource allocation problems for UEs and MTCDs can be
solved with different objectives. Consider a network with a UE set
$\mathcal{H}$ and a MTCD set $\mathcal{M}$, with each UE or MTCD
having its own utility function depending on the specific
application. When the objective of maximizing the aggregate utility
(MAX-Utility) is assumed as an example, radio resource allocation
for the eNB-to-UE and eNB-to-MTCD links can be formulated as
\begin{eqnarray}
\label{equ_max_utility} {S^*} = \arg \mathop {\max }\limits_{S \in
\mathcal{S}} \left\{ {\sum\limits_{{i} \in \mathcal{H}}
U^{H}_{i}(R^{H}_{i})+ \lambda \sum\limits_{{j} \in \mathcal{M}}
U^{M}_{j}(R^{M}_{j})} \right\}
\end{eqnarray}
\noindent where $S \in \mathcal{S}$ represents a possible resource
allocation matrix, $R^{H}_{i}$ and $R^{M}_{j}$ represent the
achievable data rate of the $i$th UE and $j$th MTCD, respectively,
$U^{H}_{i}(R^{H}_{i})$ and $U^{M}_{j}(R^{M}_{j})$ are the
corresponding utility functions, and $\lambda \in [0, 1]$ is the
unified weighting factor of M2M communication. All radio resources are
orthogonally allocated to users using the utility-based scheduling
scheme according to their pre-defined utility functions. In lieu of
the data rate, the utility has become the metric used in resource
allocation. When only slight or no utility improvement is achieved with the
increase of the data rate, users will not be assigned radio
resources.
\par

\subsection{Allocation between MTCD-to-MTCD Links}

In order to improve network efficiency, it is assumed that different
MTCD-to-MTCD links share the same radio resources. Moreover,
MTCD-to-MTCD transmission can share the resources used by the other
links owing to the low transmission power of MTCDs.  In general, the
assignment of subchannels between MTCD-to-MTCD links can be
performed through the centralized or distributed way. The former can
achieve a higher resource efficiency with much more overhead and
complexity than the latter.
\par

For the sake of implementation, a distributed \emph{graph-based}
approach can be applied for the channel assignment for MTCD-to-MTCD
links. In an interference graph, each vertex denotes a pair of
active MTCDs and an edge represents the interference condition
between two pairs. The edge exists only when the channel gain
difference between the interfering and serving links exceeds a
certain threshold. One color represents a subchannel. By measuring
the reference signal transmitted by its neighboring MTCDs, an MTCD
is able to know the device identification of each neighboring MTCD
and the pathloss between itself and its neighboring MTCD(s). Then,
each MTCD has the information of its own local interference graph
and negotiates with other MTCDs. MTCDs can request/release some
subchannels to improve on the system utility while complying with
conflict constraints imposed by other neighbors.
\par

Each vertex, i.e., an MTCD pair, is responsible for assigning its
own color, i.e., subchannel. After randomly choosing an initial
color, the vertex repeatedly chooses a color that minimizes the
number of conflicts it has with its neighbors based upon the
knowledge of its neighbors' colors. Each vertex simultaneously
chooses a color for itself. When a vertex changes the color, it also
communicates its new color to its neighbors in time. The decision to
use more than one subchannel at a vertex is probabilistic. Each
vertex determines an activation probability, dependent on its degree
and resources already occupied. Then, the vertex generates a random
number in the range of [0,1], and decides to activate one more
subchannel if the random number falls below its activation
probability. The method for determining the activation probability
can significantly impact on the performance.
\par

\section{Performance Evaluation}

In this section, the performances of LTE-A cellular networks with
M2M communications under the urban scenarios are evaluated through
system-level simulations. The detailed simulation parameters
including the channel model and system configurations are summarized
in Table~\ref{table:system:paras}~\cite{3gpp36814}, which mostly are
defined in 3rd Generation Partnership Project (3GPP) specifications.
All UEs are evenly distributed in the circular areas around each
eNB. The MTCD placement is performed as follows: 50 MTCDs are
located uniformly per sector while 50 pairs of MTCDs are deployed
uniformly at random in a floor of a building, all duty cycled at
10\%. To simulate the realistic scenarios where mixed H2H and M2M
services exist in the cellular network, different utility functions
are assumed for the UEs and MTCDs, respectively~\cite{Lee05}, i.e.,
\emph{Class} 1 for UEs while \emph{Class} 4 for MTCDs.
\par

\begin{table}[t!]
\renewcommand{\arraystretch}{1.3}
\caption{Parameter configurations in LTE-Advanced cellular networks}
\label{table:system:paras} \centering
\begin{tabular}{c|c|c}
\hline
\multicolumn{2}{c|}{\textbf{Parameter}} & \textbf{Values} \\
\hline
\multicolumn{2}{c|}{Cellular layout} & 19 cells / 3 sectors per cell \\
\hline
\multicolumn{2}{c|}{Inter-site distance (ISD)} & 500 m  \\
\hline
\multicolumn{2}{c|}{Macro UE density} & 5 UEs / sector \\
\hline
\multicolumn{2}{c|}{MTCD placement} & 5 MTCDs per sector, \\
\multicolumn{2}{c|}{} &  10 pairs of MTCDs in apartments \\
\hline
\multicolumn{2}{c|}{Macro cell shadowing standard deviation} & 8 dB \\
\hline
Macro cell shadowing correlation & Between cells & 0.5 \\
\cline{2-3} & Between sectors & 1\\
\hline
\multicolumn{2}{c|}{Max eNB transmit power} & 46 dBm\\
\hline
\multicolumn{2}{c|}{eNB antenna gain after cable loss} & 14 dBi \\
\hline
\multicolumn{2}{c|}{Max MTCD transmit power} & 14 dBm \\
\hline
\multicolumn{2}{c|}{UE antenna gain} & 0 dBi \\
\hline
\multicolumn{2}{c|}{Noise figure } & 9 dB \\
\hline
\multicolumn{2}{c|}{Apartment block} & Two stripes with \\
\multicolumn{2}{c|}{} & 1$\times$4$\times$10 (floor$\times$row$\times$column) for each stripe \\
\hline
\multicolumn{2}{c|}{Number of blocks per cell} &  1 \\
\hline
Pathloss & eNB-to-UE/MTCD & $128.1 + 37.6\log(R)$ in dB, $R$ in $km$ \\
\cline{2-3} & MTCD-to-MTCD  & LOS: $38.5+ 20\log(R), R <0.3$, $R$ in $m$   \\
        &  &  NLOS: $48.9+ 40\log(R)$, $R \geq 0.3$ \\
\hline
\end{tabular}
\end{table}

%\begin{figure}[tb]
%\begin{center}
%\includegraphics[width=4in]{CDF_Utility_Scheduling}
%\caption{Performance comparison between the Max-Utility and Proportional Fair scheduling
%schemes.} \label{fig_CDF_Utility_Scheduling}
%\end{center}
%\end{figure}

\begin{figure}[thbp]
\begin{center}
\subfigure[Effects of unified weighting
factor.]{\includegraphics[width=5.5 in]{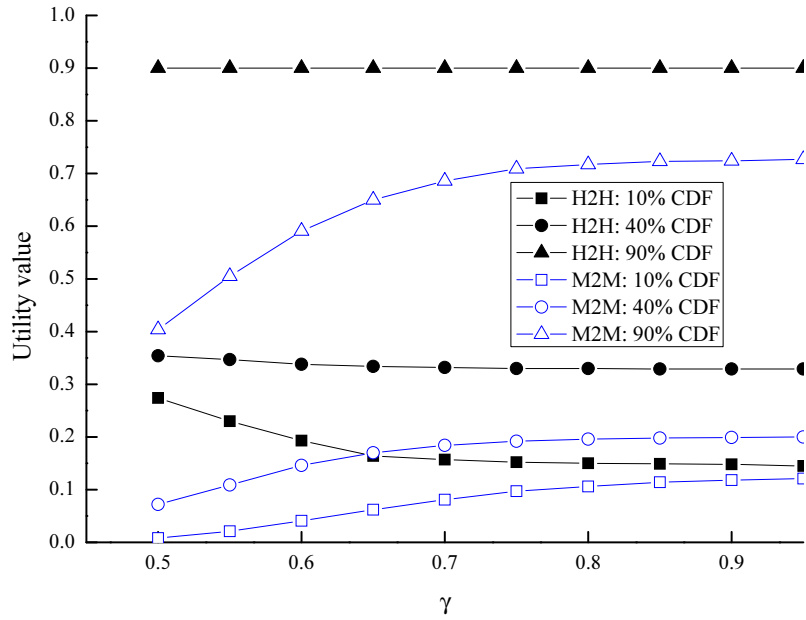}
\label{fig_CDF_Vs_alpha}} \subfigure[CDF
performances.]{\includegraphics[width=5.5
in]{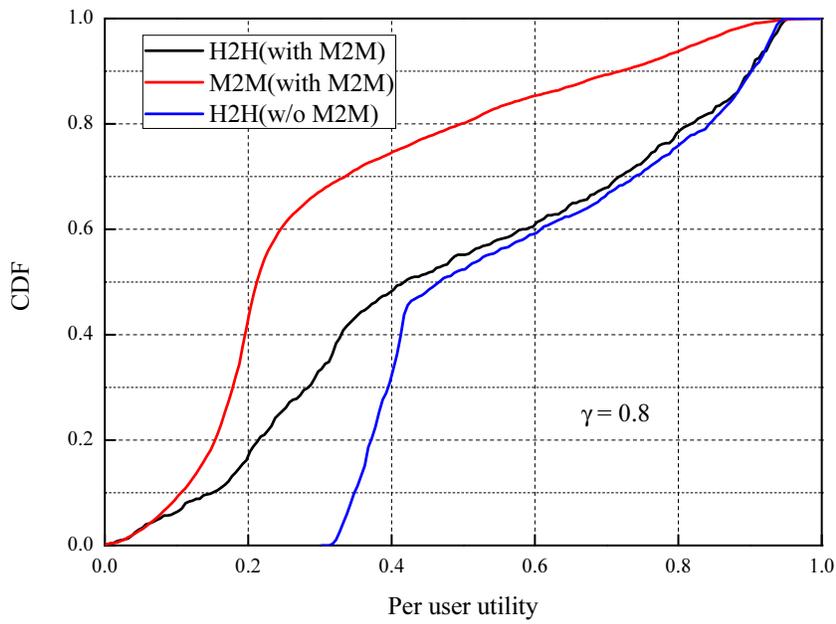}\label{fig_CDF_Utility_Scheduling}}
\caption{Performance comparison in LTE-Advanced networks
with/without M2M communication.} \label{fig_CDF_performance}
\end{center}
\end{figure}

When M2M communications are introduced into the network, the
performances of existing H2H communications are somehow affected due
to the decrease of the available radio resources. When the
\emph{Max-Utility} scheduling scheme is applied, such effects can be
controlled by adjusting the unified weighting factor of M2M
communications, i.e., $\lambda$. In Fig.~\ref{fig_CDF_Vs_alpha}, we
compare the user utility performances in terms of the given percent
point of the cumulative distribution function (CDF) in the networks
with different values of $\lambda$. With the increase of the factor
value, the performance of M2M communication is improved while the
cell edge user performance, i.e., the 10\% CDF H2H performance,
deteriorates. Such improvement in M2M communication and degradation
in H2H communication levels off when $\lambda$ is larger than 0.8.
On the other hand, the performances of H2H communication located in
the cell center, i.e., 50\% and 90\% CDF H2H performances, remain
virtually unchanged with a variation of $\lambda$. Therefore,
$\lambda=0.8$ is used for the network with mixed H2H and M2M
services. Then, Fig.~\ref{fig_CDF_Utility_Scheduling} presents the
CDF performance of the user utility with and without concurrent M2M
communications. It is observed clearly that the performance of UEs
at the cell edge is degraded when MTCDs are introduced to the
network. This is due to the parameter setting of the utility
functions for M2M and H2H communications in our simulations. The
utility of MTCDs increases more rapidly than that of the UEs in the
low rate region. Then, MTCDs rather than UEs are more likely
selected by the Max-Utility scheduling scheme when the achievable
rate is not high. In other words, we can adjust the scheduling
priority of M2M and H2H communications in the given data region by
applying the specific formats of the utility functions. On the other
hand, besides the existing H2H communications, MTCDs with
simultaneous M2M communications contribute to the aggregated cell
utility. The utility achieved by all MTCDs is larger than the
utility degradation of the cell edge UEs. Hence, the performance of
the aggregated cell utility is improved, i.e., 4.0264 with and
2.8714 without M2M communications, respectively. It is noted that
such a gain depends on several factors including the parameters
of the utility functions, the number of MTCDs, unified weighting
factor of M2M communications, etc.
\par

%Fig.~\ref{fig_CDF_Utility_Scheduling} compares the cumulative
%distribution functions (CDFs) of user utility with two different
%scheduling schemes, namely, \emph{Max-Utility} and
%\emph{Proportional Fair (PF)}. To achieve maximum system
%throughput while ensuring the proportional rate fairness among UEs
%and MTCDs, the sum of the logarithmic user throughput is maximized
%when \emph{PF} scheduling is applied. Compared to the case with the
%\emph{PF} scheduling scheme, both H2H and M2M users are able to
%achieve better performance in terms of user utility using the
%\emph{Max-Utility} scheduling scheme. It should be noted that there
%are several steps in the M2M curve with \emph{Max-Utility}
%scheduling. In the given rate region, i.e., the rates between
%14 and 310, since the user utility of \emph{Class} 2 is much higher
%than that of \emph{Class} 1, users of \emph{Class} 2 are always
%selected by the \emph{Max-Utility} scheduling scheme, while the
%\emph{PF} scheduling scheme does not. Moreover, there are four users
%of \emph{Class} 2 in our simulations so that the number of steps in
%the curve is four.\par

\begin{figure}[thbp]
\begin{center}
\includegraphics[width=6.5in]{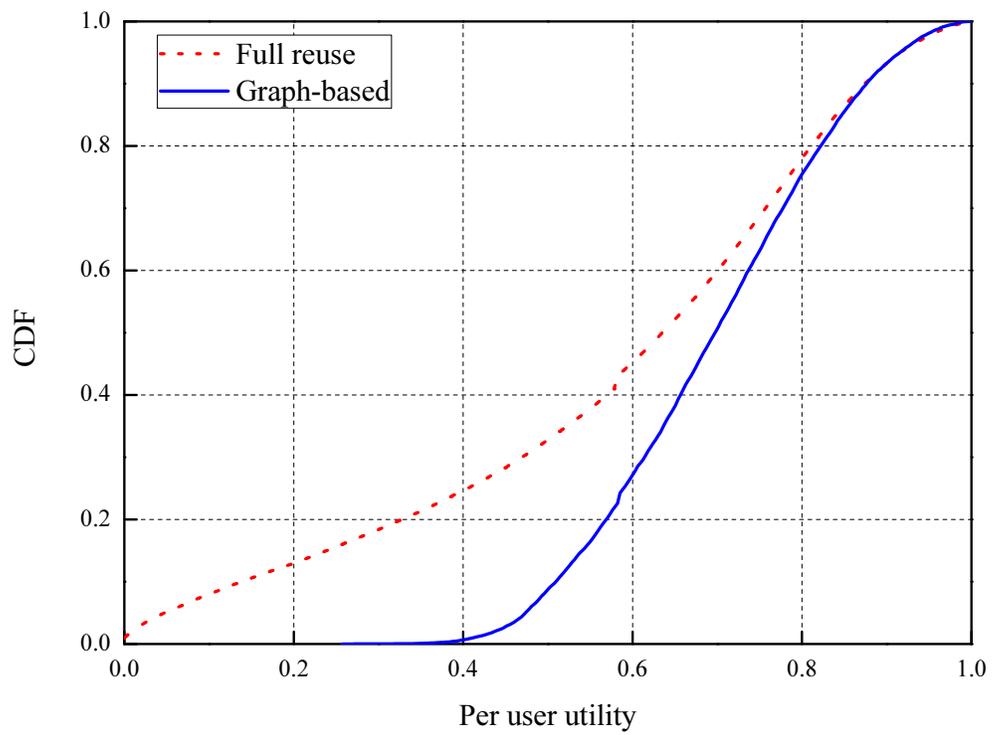}
\caption{Performance comparison between the full reuse and graph-based channel allocation schemes
for MTCDs with peer-to-peer transmission.}
\label{fig_CDF_Utility_Graph_IC}
\end{center}
\end{figure}

As shown in Fig.~\ref{fig_CDF_Utility_Graph_IC}, we examine the user
utility performance of MTCDs deployed in urban buildings with
peer-to-peer transmission, where the full reuse and
\emph{graph-based} channel allocation approaches are applied,
respectively. When the graph-based approach is applied to deal with
radio resource allocation between MTCDs, the interference between
the different links is well controlled. Then, compared with the full
reuse approach, the received SINRs of MTCDs are increased especially
at the cell edge. Consequently, the utility performance is improved,
e.g., in 90\% operational cases, the full reuse approach achieves a
utility of only 0.15, whereas the graph-based approach yields at
least 0.5. Moreover, the gain in the low or medium data rate region
is more obvious than that in the high data rate region. This is
because the utility of MTCDs is a non-linear function of the data
rate per resource block, i.e., the utility increases much more
rapidly in the low or medium data rate region than in the high data
rate region.\par

\section{Conclusions}

M2M communications are clearly an emerging technology and a
facilitator of the IoT by means of, among others, cellular
technology. It has thus gained increasing attention in LTE-Advanced
cellular network designs. In this paper, we first presented the
required network architectural enhancements with the introduction of
various transmission schemes related to MTCDs. Then, several radio
resource allocation schemes for different transmission links have
been proposed with the aim of minimizing co-channel interference and
maximizing network efficiency. Our simulation results demonstrate
that the proposed schemes can improve the network performance in
terms of user utility. \par

In the next step, practical issues will be paid more attention to
when designing  new resource allocation schemes for M2M
communications. Firstly, typical application scenarios and M2M
service features will be kept in line with the development of the
standardization bodied such as 3GPP and ETSI M2M. Then, to meet the specified
requirements of given scenarios and services, more types of schemes
are developed with different objectives. Moreover, the overhead
and complexity for implementing the schemes have to be considered in
order to strike a right balance between  performance and cost. It is
expected that the well-designed resource allocation schemes can
bring to operators remarkable benefits at affordable costs in
LTE-Advanced M2M-enabled cellular networks.\par

\section*{Acknowledgment}

This work was supported in part by ICT project INFSO-ICT-258512
EXALTED, National Key Technology R\&D Program of China under Grant
2009ZX03003-008-01 and Research Fund for the Doctoral Program of
Higher Education under Grant 20090005120002.
\par

\section*{Biography}

\begin{small}

\textbf{Kan Zheng}
(M'03, SM'09) received the B.S., M.S. and Ph.D degree from Beijing
University of Posts\&Telecommunications (BUPT), China, in 1996, 2000 and
2005, respectively, where he is currently associate professor. He
worked as a senior researcher in the companies including Siemens,
Orange Labs R\&D (Beijing), China. His current research interests lie
in the field of machine-to-machine (M2M) communication, cooperative
communication and heterogeneous networks.

\textbf{Fanglong Hu}
received the B.S. degree from Beijing
University of Posts\&Telecommunications (BUPT), China, in 2009. Since then
he has been working toward a M.S. degree in BUPT. His research interests
include machine-to-machine (M2M) communication and heterogeneous networks.

\textbf{Wei Xiang} (M'04, SM'10) received the B.Eng. and M.Eng.
degrees, both in electronic engineering, from the University of
Electronic Science and Technology of China, Chengdu, China, in 1997
and 2000, respectively, and the Ph.D. degree in telecommunications
engineering from the University of South Australia, Adelaide,
Australia, in 2004. Since January 2004, he has been with the Faculty
of Engineering and Surveying, University of Southern Queensland,
Toowoomba, Australia, where he was first an Associate Lecturer in
Computer Systems Engineering from 2004 to 2006, then a Lecturer from
2007 to 2008, and currently holds a faculty post of Senior Lecturer.
His research interests are in the broad area of communications and
information theory, particularly coding and signal processing for
multimedia communications systems. He was a visiting scholar to
Nanyang Technological University from January to June 2008, and the
University of Mississippi from October 2010 to March 2011,
respectively. He received a prestigous Queensland International
Fellowship awarded by the State Government of Queensland,
Commonwealth of Australia, in 2010.

\textbf{Mischa Dohler}
 (mischa.dohler@cttc.es) is now leading Intelligent
Energy [IQe] at CTTC in Barcelona, as well as being CTO of
Worldsensing. Prior to this, he was at France Telecom and King¡¯s
College London. He is working on smart grid, machine-to-machine,
femto, cooperative, cognitive and docitive networks. He has
published several books and more than 130 refereed papers, holds
several patents, has given numerous tutorials, and participated in
standardization activities. He has been chair and TPC of various
conferences. He is and has been editor for numerous IEEE and
non-IEEE journals, as well as being the EiC of ETT. He is Senior
Member of the IEEE and fluent in 6 languages.

\textbf{Wenbo Wang} (M'98) received his B.S., M.S. and Ph.D degree
from Beijing University of Posts\&Telecommunications (BUPT), China,
in 1986, 1989 and 1992 respectively. He is currently a professor and
dean of graduate school in BUPT. His research interests include
signal processing, mobile communications and wireless network.

\end{small}

\newpage

\newpage

\newpage

\newpage

\newpage

\newpage

\end{document}